\documentclass[conference]{IEEEtran}
\usepackage{cite}
\usepackage{amsmath,amssymb,amsfonts}
\usepackage{algorithmic}
\usepackage{graphicx}
\usepackage{textcomp}
\usepackage{xcolor}
\usepackage{booktabs}
\usepackage{adjustbox}
\usepackage{makecell}
\usepackage{array}
\usepackage{multirow}
\usepackage{caption}
\def\BibTeX{{\rm B\kern-.05em{\sc i\kern-.025em b}\kern-.08em
    T\kern-.1667em\lower.7ex\hbox{E}\kern-.125emX}}
\begin{document}

\title{Multilingual Dataset Integration Strategies for Robust Audio Deepfake Detection: A SAFE Challenge System\\
}




\author{\IEEEauthorblockN{Hashim Ali, Surya Subramani, Lekha Bollinani, Nithin Sai Adupa, Sali El-Loh, Hafiz Malik\thanks{Corresponding author: hafizm@umich.edu}}
\IEEEauthorblockA{\textit{Electrical and Computer Engineering} \\
\textit{University of Michigan}\\
Dearborn, USA}
}


\maketitle

\begin{abstract}

The SAFE Challenge evaluates synthetic speech detection across three tasks: unmodified audio, processed audio with compression artifacts, and laundered audio designed to evade detection. We systematically explore self-supervised learning (SSL) front-ends, training data compositions, and audio length configurations for robust deepfake detection. Our AASIST-based approach incorporates WavLM large frontend with RawBoost augmentation, trained on a multilingual dataset of 256,600 samples spanning 9 languages and over 70 TTS systems from CodecFake, MLAAD v5, SpoofCeleb, Famous Figures, and MAILABS. Through extensive experimentation with different SSL front-ends, three training data versions, and two audio lengths, we achieved second place in both Task 1 (unmodified audio detection) and Task 3 (laundered audio detection), demonstrating strong generalization and robustness.

\end{abstract}


\section{Introduction}
The rapid advancement of text-to-speech (TTS) synthesis technologies has created an urgent need for robust audio deepfake detection systems. Although these technologies offer beneficial applications, their misuse for creating convincing fake audio poses significant security and societal risks. The SAFE (Synthetic Audio Forensics Evaluation) Challenge\footnote{https://stresearch.github.io/SAFE/}  represents a critical step forward in addressing these challenges by providing a comprehensive evaluation framework that emphasizes real-world applicability and generalizability across diverse audio sources. Unlike traditional evaluation protocols that often focus on controlled laboratory conditions, the SAFE Challenge is designed to test detection systems in realistic scenarios where spoofing attacks may originate from unknown synthesis methods, undergo various processing operations, or be deliberately laundered to evade detection. This emphasis on generalizability across diverse sources reflects the practical challenges faced by deepfake detection systems deployed in real-world environments, where the characteristics of spoofed audio can vary dramatically from training data.

A significant limitation of current deepfake detection research is the reliance on single, often clean datasets for training. Most existing systems are trained on well-curated datasets such as ASVspoof \cite{nautsch2021asvspoof, wang2020asvspoof, kamble2020advances}, which, while valuable for controlled evaluation, may not adequately prepare models for the diversity of spoofing techniques and audio characteristics encountered in practice. This training paradigm can lead to models that perform well on in-domain test sets but fail to generalize to new attack types, different languages, or varying audio quality conditions.

To address this generalization challenge, we conducted four iterative experiments that systematically explored dataset integration strategies for training robust audio deepfake detection systems. Our approach progressively incorporated multiple datasets that span different languages, spoofing techniques, and audio quality conditions, while also investigating the impact of factors such as audio segment length and SSL (self-supervised learning) front-end model selection. These experiments were designed to understand how diverse multilingual datasets can improve the robustness of detection systems when faced with unknown spoofing attacks. We evaluated our approach on both the SAFE Challenge tasks and the In-The-Wild (ITW) dataset \cite{muller_does_2022}, a community-standard benchmark to assess generalized performance of audio deepfake detection systems.

Our work makes several practical contributions to the audio deepfake detection community. First, we provide an empirical evaluation of how the combination of datasets affects detection performance across the three SAFE Challenge tasks and on the ITW benchmark \cite{muller_does_2022}. Second, we analyze performance patterns that reveal insights into task-specific challenges, particularly the difficulty of detecting laundered audio. Third, we demonstrate the value of incorporating diverse multilingual training data for improving cross-domain generalization capabilities. Finally, we performed a source-level analysis for both generated and pristine (authentic) sources from the SAFE Challenge, revealing critical vulnerabilities and failure patterns across different synthesis methods and audio processing scenarios. These findings offer actionable insights for researchers and practitioners working to develop more robust audio deepfake detection systems.

The remainder of this paper is organized as follows. Section \ref{related} reviews existing audio deepfake detection datasets and self-supervised learning approaches relevant to spoofing detection. Section \ref{safe} describes the SAFE Challenge framework and evaluation protocol. Section \ref{system} presents our model architecture and SSL front-end selection methodology. Section \ref{dataset_integration} details our dataset integration strategies and systematic experimental design. Section \ref{training} covers training configurations and implementation details. Section \ref{res} provides comprehensive results and analysis from both SAFE Challenge and ITW evaluations. Finally, Section \ref{conc} concludes with implications for the deepfake detection community.


\section{Related Work} \label{related}

\subsection{Audio Deepfake Detection Datasets}
The development of robust audio deepfake detection systems has been driven by the availability of diverse evaluation datasets, each addressing specific aspects of the spoofing detection challenge.

\textbf{ASVspoof Series:} The ASVspoof Challenges \cite{wang2020asvspoof, liu_asvspoof_2023, yamagishi2021asvspoof, wang24_asvspoof} have provided foundational datasets for the community. ASVspoof 2015 \cite{wu2014asvspoof} introduced the first spoofing database with 10 TTS systems. ASVspoof 2019 \cite{wang2020asvspoof} LA expanded this with 19 TTS systems (6 known, 13 unknown), including WaveNet, Tacotron2, and traditional approaches, establishing the standard for controlled evaluation. ASVspoof 2021 \cite{yamagishi2021asvspoof} introduced two tracks: Logical Access (LA) with 13 systems and Deepfake (DF) with more than 100 attack methods, significantly expanding the attack diversity. The recent ASVspoof 5 \cite{wang24_asvspoof} represents the most complex design with multilingual data and a wide variety of attacks on a large scale.

\textbf{Multilingual and Cross-Domain Datasets:} MLAAD (Multi-Language Audio Anti-Spoofing Dataset) \cite{muller2024mlaad} addresses linguistic diversity with 91 TTS systems across 38 languages, providing 420.7 hours of synthetic speech from 42 different architectures. This dataset specifically targets the language bias present in predominantly English-focused datasets. MLAAD dataset only provides the synthetic audio samples. The M-AILABS Speech Dataset\footnote{https://github.com/imdatceleste/m-ailabs-dataset} complements this by providing authentic multilingual speech samples across multiple languages, sourced from audiobooks and public figures speeches.

\textbf{Real-World and Noisy Conditions:} SpoofCeleb \cite{jung_spoofceleb_2024} leverages VoxCeleb1 \cite{zhang_beijing_2021} as source data, training 23 contemporary TTS systems in real-world noisy conditions to bridge the gap between clean laboratory data and practical applications. In-The-Wild (ITW) \cite{muller_does_2022} provides audio deepfakes collected from social media platforms, offering genuine ``in-the-wild'' evaluation conditions. Hashim et al. \cite{ali2024audio} applied various real-world processing on ASVSpoof19 audio to generate laundered version of the dataset.

\textbf{Technology-Specific Datasets:} CodecFake \cite{xie2024codecfake} introduces the first dataset focused specifically on neural codec-based synthesis, featuring 15 codec models from 6 frameworks, including SpeechTokenizer \cite{zhang2024speechtokenizer}, Encodec \cite{defossez2022high}, and novel approaches like FunCodec \cite{du2024funcodec}. The DFADD dataset \cite{du2024dfadd} focuses on diffusion- and flow-matching-based TTS systems, including GradTTS \cite{popov2021grad}, NaturalSpeech2 \cite{shen2023naturalspeech}, Style-TTS2 \cite{li2024styletts}, Matcha-TTS \cite{mehta2024matcha}, PFlow-TTS \cite{kim2024p}, etc.

\textbf{Famous Figures Dataset \cite{ali2025collecting}:} 
Motivated by recent incidents such as the Biden robocall \cite{elliott_biden_2024}, and fabricated recordings of London Mayor Sadiq Khan making inflammatory remarks \cite{Marianna_sadiq_2024}, we curated a specialized dataset for protecting famous figures from voice cloning attacks. This dataset was developed by collecting high-quality bonafide
speech samples of famous figures from YouTube, and then generating the corresponding synthetic speech using various TTS approaches. This dataset provides a deep coverage of political personalities using cutting-edge 2024-2025 TTS systems, such as StyleTTS2 \cite{li2024styletts}, XTTSv2 \cite{casanova2024xtts}, F5TTS \cite{chen2024f5}, E2TTS \cite{eskimez2024e2}, etc.

\subsection{Self-Supervised Learning in Audio Spoofing Detection}

Self-supervised learning (SSL) has emerged as a powerful approach in audio deepfake detection, addressing challenges such as limited labeled data, generalization to unseen attacks, and robustness across domains. SSL leverages large amounts of unlabeled audio to learn discriminative representations, which can then be fine-tuned or used directly for downstream deepfake detection tasks. Tak et al. \cite{tak2022rawboost} investigated the use of Wav2Vec2 for audio spoof detection, demonstrating that SSL-based models achieved SOTA performance even when trained exclusively on bona fide speech samples. They also introduced RawBoost, a data augmentation framework to enhance robustness against real-world distortions.

The cross-dataset generalization capabilities of SSL models have been particularly important. Pascu et al. \cite{pascu2024towards} showed that using frozen SSL representations with simple classifiers significantly improved performance, reducing the Equal Error Rate (EER) from 30.9\% to 8.8\% across eight deepfake datasets,  while emphasizing the importance of model calibration for reliable confidence scores. Recent works \cite{stourbe2024exploring, combei2024wavlm} have demonstrated that different SSL models capture complementary spoofing artifacts, with WavLM consistently outperforming other models. Stourbe et al. and Combei et al. showed that ensemble approaches combining multiple SSL variants with different backend architectures can achieve superior results through late fusion techniques.

\begin{figure*}[htbp]
\centering
\begin{minipage}[c]{0.48\textwidth}
    \centering
    \includegraphics[width=\textwidth]{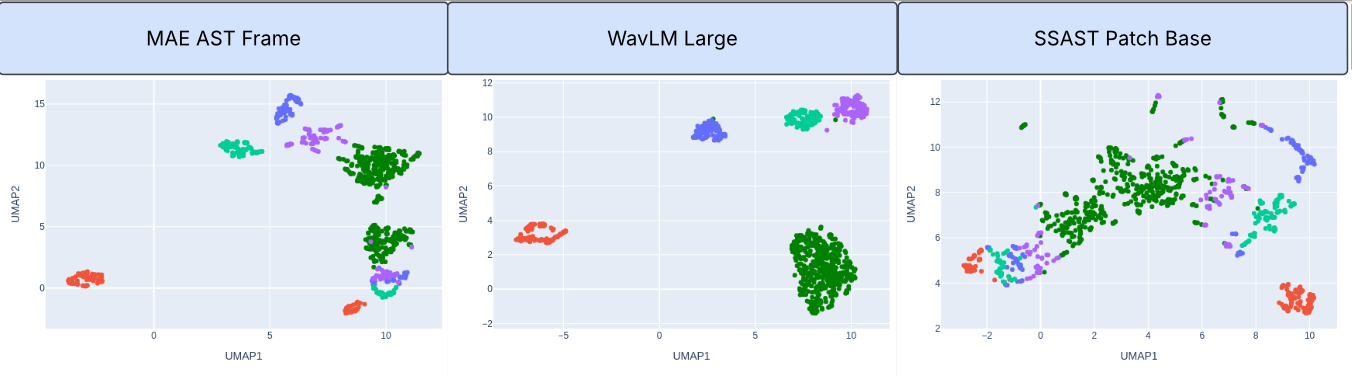}
    \vspace{0.1cm}
    \includegraphics[width=\textwidth]{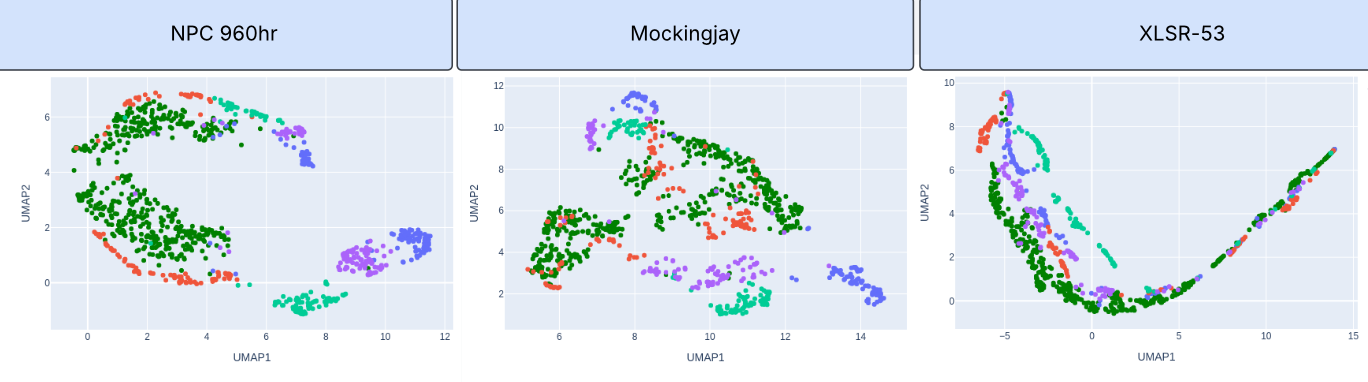}
    \vspace{-0.3cm}
    \setlength{\abovecaptionskip}{0pt}
    \caption{UMAP visualizations of SSL model representations for ASVSpoof 2019 LA eval data. Top: Comparison of MAE AST Frame, WavLM Large, and SSAST Patch Base models. Bottom: Additional SSL model comparisons including NPC 960hr, Mockingjay, and XLSR-53. Green color represent bonafide, other colors represent various deepfakes.}
    \label{fig:umap_ssl_models}
\end{minipage}
\hfill
\begin{minipage}[c]{0.48\textwidth}
    \centering
    \scriptsize
    \captionof{table}{SAFE Challenge Audio Sources by Task}
    \renewcommand{\arraystretch}{1.2} 
    \begin{tabular}{|>{\centering\arraybackslash}m{1.5cm}|
                >{\centering\arraybackslash}m{1.8cm}|
                p{4cm}|}
    \hline
    \textbf{Task} & \textbf{Category} & \textbf{Source Names} \\
    \hline
    \multirow{14}{*}{\shortstack{Task 1\\(Unmodified)}} 
     & \multirow{10}{*}{Real Audio} & Mandarin Podcast, FLEURS German, VSP Semi-professional, YouTube phonecall, VSP Documentary, Arabic Speech Corpus, High Quality Podcasts, Japanese Shortwave, Conference, English Podcast, FLEURS English, Djeco, Digitized Cassette, Librivox, Old Radio, phone home, Russian Audiobook, VSP Home Mic, Radio Drama, VSP Professional \\
    \cline{2-3}
     & \multirow{3}{*}{TTS Systems} & elevenlabs, fish, hierspeech, kokoro, parler, seamless, style, cartesia, edge, f5, metavoice, openai, zonos \\
    \hline
    \multirow{6}{*}{Task 2} & \multirow{6}{*}{Processed Audio} & aac 16k, encodec, focalcodec, mp3-aac-mp3 16k, mp3-aac 16k, mp3 16k, mp3 VBR, noise, opus 16k, phone audio, pitch shift, resample down/up, semanticodec, snac, speech filter, time stretch, vorbis 16k \\
    \hline
    Task 3 & Laundered Audio & car, played, played reverb car, reverb \\
    \hline
    \end{tabular}
    \label{tab:safe_sources}
\end{minipage}
\end{figure*}

\section{SAFE Challenge Overview} \label{safe}
The SAFE Challenge evaluates \cite{kirill2025safe} audio deepfake detection systems across three distinct tasks designed to test different aspects of robustness.

\begin{enumerate}
    \item \textbf{Task 1 (Generated Audio):} Detection of unmodified synthetic speech directly from TTS model output, testing basic spoofing detection capabilities.

    \item \textbf{Task 2 (Processed Audio):} Detection of generated samples that have undergone compression and resampling operations, simulating real-world distribution scenarios.

    \item \textbf{Task 3 (Laundered Audio):} Detection of deliberately processed synthetic audio designed to avoid detection systems, representing adversarial laundering attacks.
\end{enumerate}

The SAFE evaluation dataset comprises human- and machine-generated speech audio tracks with several key characteristics that emphasize real-world applicability. Human speech samples are sourced from multiple origins and languages, ranging from high-quality studio recordings to lower-quality in-the-wild online recordings, as detailed in Table \ref{tab:safe_sources} under Task 1 - Real Audio sources. Machine-generated samples are created using state-of-the-art TTS models, including both open-source and closed-source systems such as ElevenLabs, OpenAI, and various neural synthesis approaches (Table \ref{tab:safe_sources}, Task 1 - TTS Systems). Beyond unmodified audio detection, the challenge incorporates realistic degradation scenarios through Task 2, which applies various processing operations including compression codecs (AAC, MP3, Opus), resampling, pitch shifting, and additive noise to simulate real-world distribution conditions (Table \ref{tab:safe_sources}, Task 2 - Processed Audio). Task 3 addresses adversarial laundering attacks that deliberately process synthetic audio to evade detection systems, employing techniques such as playback through car environments, acoustic reverberation, and re-recording methods (Table \ref{tab:safe_sources}, Task 3 - Laundered Audio). Audio files vary in length up to 60 seconds with diverse compression formats, and the dataset maintains balance across different sources. Critically, the competition employs a fully blind evaluation protocol where no training data is released, ensuring that the systems must generalize from external training data to unknown test conditions.

\section{Model Architecture and SSL Selection} \label{system}
Our detection system employs a two-stage architecture combining self-supervised learning (SSL) front-ends with the AASIST (Audio Anti-Spoofing using Integrated Spectro-Temporal graph attention networks) back-end as proposed by \cite{tak2022automatic}. To select optimal SSL models, we conducted UMAP visualization analysis on various SSL front-ends using ASVSpoof 2019 LA database. Figure \ref{fig:umap_ssl_models} shows the UMAP visualization of MAE-AST Frame \cite{baade2022mae}, WavLM Large \cite{Chen2021WavLM}, SSAST Patch Base \cite{gong2022ssast}, NPC 960hr \cite{liu2020non}, Mockingjay \cite{encoders2020mockingjay}, and XLSR-53 models \cite{babu2021xls}. This analysis revealed that WavLM Large and MAE-AST Frame models provided the most discriminative feature representations for distinguishing between authentic and synthetic audio. The AASIST backend was chosen for its proven effectiveness in modeling both spectral and temporal spoofing artifacts.

\section{Dataset Integration Strategy} \label{dataset_integration}

\subsection{TTS System Coverage Analysis} \label{tts_coverage}

\begin{table*}[t]
\centering
\caption{TTS Systems Coverage Across Deepfake Detection Datasets}
\label{tab:tts_coverage}
\begin{tabular}{|>{\centering\arraybackslash}m{2.5cm}|>{\centering\arraybackslash}m{3cm}|>{\centering\arraybackslash}m{4.5cm}|>{\centering\arraybackslash}m{4cm}|}
\hline
\textbf{Dataset} & \textbf{Total TTS Systems} & \textbf{Key TTS Architectures} & \textbf{Complementary Features} \\
\hline
CodecFake & 15 codec models (6 frameworks) & SpeechTokenizer, AcademicCodec, AudioDec, Encodec & Neural audio codec focus \\
\hline
ASVspoof 2019 LA & 19 systems (6 known + 13 unknown) & WaveNet, Tacotron2, VAE-based, GMM-UBM & Traditional + neural mix \\
\hline
MLAAD & 54-91 systems (21-42 architectures) & VITS variants, SpeechT5, multilingual models & Multilingual diversity \\
\hline
SpoofCeleb & 23 systems & Contemporary TTS on VoxCeleb1 & Real-world noisy conditions \\
\hline
Famous Figures & 10 systems & StyleTTS2, XTTSv2, F5TTS, E2TTS, FishSpeech & Real-world, Latest 2024-2025 TTS \\
\hline
\end{tabular}
\end{table*}

To systematically identify complementary datasets for optimal generalization performance, we conducted a comprehensive analysis of TTS system coverage across existing deepfake detection datasets. This analysis guides our strategic dataset selection approach. Table \ref{tab:tts_coverage} provides a brief coverage of the TTS systems available in datasets used by this study.

\subsubsection{Complementarity Analysis}
Our analysis revealed three key insights: (1) \textbf{Technology Coverage Gaps} - most datasets focus on specific paradigms, with Famous Figures uniquely covering 2024-2025 TTS systems absent elsewhere; (2) \textbf{Optimal Combinations} - highest complementarity achieved by combining codec-based (CodecFake), multilingual (MLAAD), real-world noisy (SpoofCeleb), and cutting-edge (Famous Figures) approaches; (3) \textbf{Acoustic Condition Diversity} - combining clean studio, multilingual audiobook, and real-world noisy data provides comprehensive acoustic coverage.

\subsubsection{Strategic Implications}
This analysis demonstrates that systematic dataset combination based on complementary TTS coverage can address generalization challenges more effectively than single-dataset approaches, providing the foundation for our iterative experimental design.

\subsection{Iterative Experimental Design}
We systematically explored dataset integration through four iterative experiments, each building on the insights from the previous iterations. Our approach was motivated by the hypothesis that diverse multilingual datasets are crucial for generalization to unknown spoofing attacks.

\subsubsection{Iteration 1 (Baseline)}
We established a baseline using only the ASVspoof 2019 LA training dataset, which comprises 25,380 samples (2,580 real, 22,800 fake). This single-dataset approach represents the conventional training paradigm and provided a reference point for measuring the impact of dataset diversification.

\subsubsection{Iteration 2 (Multi-dataset Integration)}
We expanded training data to include five complementary datasets with strategic sampling from each. For all datasets except ASVspoof 2019 LA, we applied an 80-20 split for training and validation data allocation.

\begin{itemize}
    \item \textbf{ASVspoof 2019 LA:} Complete dataset with 25,380 samples providing established spoofing detection benchmarks
    
    \item \textbf{M-AILABS:} We sampled 20,000 multilingual authentic speech samples spanning 8 languages (English, French, German, Italian, Polish, Russian, Spanish, and Ukrainian).  Following the 80-20 split, these samples were divided into 16,000 for training and 4,000 for validation, providing authentic multilingual speech baselines.
    
    \item \textbf{MLAAD (Multi-Language Audio Anti-Spoofing Dataset):} We sampled 47,200 multilingual synthetic samples focusing specifically on languages available in M-AILABS plus Hindi language. This strategic language selection ensured consistency with our authentic multilingual data while covering diverse TTS architectures across multiple linguistic contexts.

    \item \textbf{CodecFake A2:} Since we were only interested in TTS systems, we specifically included only the A2 subset which uses VALL-E X, a modern neural codec-based TTS approach. We utilized 7,109 samples for training and 1,778 samples for validation, representing contemporary codec-based synthesis methods.

    \item We sampled 12,734 custom samples with 7,200 authentic and 5,534 synthetic samples from our Famous Figures dataset \cite{ali2025collecting}. This dataset provides latest TTS systems for specific high-profile speakers.
\end{itemize}

This resulted in 108,423 training samples (25,780 authentic, 82,643 synthetic) and 45,606 validation samples, significantly expanding both linguistic and technical diversity compared to the single-dataset baseline.

\subsubsection{Iteration 3 (Audio Length Optimization)}
Using the same multi-dataset composition from Iteration 2, we increased audio segment length from 4 seconds to 12 seconds based on the observation that SAFE audio files can extend up to 60 seconds. This change was motivated by the hypothesis that longer temporal context would improve detection of complex spoofing artifacts that may require extended analysis windows.

\subsection{Iteration 4 (Strategic Integration)}
Based on our TTS system coverage analysis from Section \ref{tts_coverage}, which identified optimal complementarity combinations for maximum technological and acoustic diversity, we refined the dataset composition through strategic optimization. This iteration specifically implements the findings that combining codec-based synthesis, multilingual diversity, real-world noisy conditions, and cutting-edge TTS technology provides the most comprehensive training foundation. The focus of Iteration 4 includes achieving balanced real/fake distribution, incorporating unseen languages and TTS systems, and ensuring comprehensive coverage across different synthesis paradigms.

\textbf{SpoofCeleb Integration:} Added 100,000 training samples (50,000 authentic, 50,000 synthetic) and 20,000 validation samples from this real-world noisy dataset, providing exposure to practical audio conditions that bridge the gap between clean laboratory data and real-world deployment scenarios.

\textbf{MLAAD Language Selection:} Strategically sampled 60,000 samples focusing on languages present in M-AILABS (English, German, Spanish, French, Italian, Polish, Russian, Ukrainian) plus Hindi. We implemented a careful train/validation split strategy to ensure both known and unknown TTS system coverage within each language:

\begin{itemize}
    \item \textbf{English:} From 36 TTS systems, we randomly selected 7 systems for validation (2,000 samples) and used the remaining 29 systems for training (7,000 samples).

    \item \textbf{Ukrainian:} Used entirely for validation (5,000 samples), providing a completely unknown language condition for language generalization.

    \item \textbf{German, Spanish, French, and Italian:} For each language, we randomly selected 2 TTS systems for validation (2,000 samples each) with remaining systems allocated to training (7,000 samples for German, 6,000 samples each for Spanish, French, and Italian).

    \item \textbf{Polish:} Used 1 TTS system for validation (1000 samples) with remaining systems for training (5000 samples).

    \item \textbf{Russian:} Used entirely for training (5,000 samples), providing additional training diversity.
    
    \item \textbf{Hindi:} Dedicated entirely to training (2,000 samples), expanding linguistic coverage beyond European languages.

\end{itemize}

This careful partitioning resulted in 44,000 training samples and 16,000 validation samples, with validation splits containing both known architectures from unknown speakers and completely unknown language conditions (Ukrainian).

\textbf{Famous Figures Refinement:} Focused on 16,000 samples with balanced authentic/synthetic distribution, emphasizing Donald Trump and JD Vance for both categories while using other speakers only for authentic samples. This dataset uniquely covers 2024-2025 TTS systems, which are absent from other datasets.

\textbf{CodecFake Integration:} Maintained 6,500 training and 1,600 validation samples from VALL-E X system to preserve codec-based synthesis representation.

\textbf{M-AILABS Expansion:} Increased to 60,000 samples (44,000 training, 16,000 validation) to better balance authentic multilingual content with the expanded synthetic data.

The final dataset comprised 200,000 training samples (101,200 authentic, 99,600 synthetic) and 56,600 validation samples (29,200 authentic, 27,400 synthetic), representing the most comprehensive multilingual, multi-domain training configuration guided by our systematic complementarity analysis. This configuration successfully achieves the targeted balanced real/fake distribution (approximately 50-50 split), incorporates multiple unseen languages and unknown TTS systems in validation splits, and provides comprehensive coverage across traditional neural TTS, codec-based synthesis, multilingual approaches, and cutting-edge 2024-2025 technologies.

\section{Training Details} \label{training}
All audio samples were resampled to 16 kHz and padded or cropped to fixed lengths depending on the iteration: 4 seconds for Iterations 1-2 and 12 seconds for Iterations 3-4. The longer audio segments in later iterations were chosen to provide extended temporal context for detecting complex spoofing artifacts, given that SAFE challenge audio files can extend up to 60 seconds.

We applied RawBoost data augmentation \cite{tak2022rawboost} across all iterations, using linear and non-linear convolutive noise combined with impulsive signal-dependent additive noise strategies optimal for logical access scenarios. All models were trained using Adam optimizer with learning rate of $10^{-6}$, and binary cross-entropy loss with class weighting. We used S3PRL toolkit \cite{yang21c_interspeech, yang2024large} to extract SSL embeddings: WavLM (1024 dimensions) and MAE-AST Frame (768 dimensions), both fed to 128-dimensional fully connected layers before AASIST back-end classification. Training was conducted separately for each iteration over 50 epochs on Nvidia A100 GPU, with reproducible results available through open source code\footnote{
https://github.com/issflab/ssl-antispoofing/}.

\section{Results and Analysis} \label{res}

Table \ref{tab:results} presents the performance progression across our four iterative experiments on both the SAFE Challenge tasks and the ITW benchmark. The results demonstrate the systematic impact of dataset integration strategies on detection performance across different evaluation scenarios.

\textbf{Iteration 1 (Baseline):} Using traditional AASIST trained solely on ASVspoof 2019 LA with 4-second audio segments, we achieved baseline performance of 0.531, 0.589, and 0.492 balanced accuracy on Tasks 1, 2, and 3 respectively.

\begin{table}[t]
\centering
\caption{Performance Progression Across Four Iterations on SAFE Challenge Tasks and ITW Benchmark.}
\label{tab:results}
\begin{tabular}{|c|c|ccc|cc|}
\hline
\multirow{2}{*}{Iter} & \multirow{2}{*}{SSL Model} & \multicolumn{3}{c|}{SAFE Challenge (BA)} & \multicolumn{2}{c|}{ITW Benchmark} \\
\cline{3-7}
& & Task 1 & Task 2 & Task 3 & BA & EER (\%) \\
\hline
1 & AASIST & 0.531 & 0.589 & 0.492 & 0.616 & 35.61 \\
2 & WavLM & 0.745 & 0.587 & 0.478 & 0.875 & 8.46 \\
2 & MAE-AST & 0.607 & 0.587 & 0.597 & 0.648 & 24.79 \\
3 & WavLM & 0.766 & 0.765 & 0.518 & 0.856 & 12.05 \\
4 & WavLM & \textbf{0.810} & \textbf{0.819} & 0.496 & \textbf{0.905} & \textbf{8.42} \\
4 & MAE-AST & 0.640 & 0.536 & \textbf{0.623} & 0.603 & 39.9 \\
\hline
\end{tabular}
\footnotesize
\end{table}

\textbf{Iteration 2 (Multi-dataset Integration):} Incorporating five complementary datasets (ASVspoof 2019 LA, M-AILABS, MLAAD, CodecFake A2, and Famous Figures) with SSL front-ends yielded substantial improvements. WavLM Large achieved 0.745, 0.587, and 0.478 on Tasks 1, 2, and 3, representing a 40.3\% improvement on Task 1. MAE-AST Frame showed 0.607, 0.587, and 0.597, demonstrating competitive performance on Task 3.

\textbf{Iteration 3 (Audio Length Optimization):} Extending audio segments from 4 to 12 seconds using the same multi-dataset composition produced notable gains for WavLM Large: 0.766 (Task 1), 0.765 (Task 2), and 0.518 (Task 3). The 30.3\% improvement in Task 2 performance specifically validates the importance of longer temporal context for detecting processed audio artifacts.

\textbf{Iteration 4 (Strategic Integration):} Our final configuration, incorporating SpoofCeleb, refined dataset balancing and strategic train-validation splits, achieved optimal performance with WavLM Large: 0.810 (Task 1), 0.819 (Task 2), and 0.496 (Task 3). This represents cumulative improvements of 52.5\% and 39.0\% for Tasks 1 and 2 respectively from the baseline. Task 3 performance remained stable around 0.49-0.52 across iterations, reflecting the inherent challenge of detecting adversarially laundered audio.

\subsection{SAFE Challenge Performance}

Our best configuration (Iteration 4, WavLM Large) demonstrated strong performance across both evaluation phases of the SAFE Challenge. On the public leaderboard, we achieved second place across all three tasks (Tasks 1, 2, and 3). On the private leaderboard, which is a superset of the public split, we secured third place for all tasks. This consistent top-tier ranking among international research teams validates the effectiveness of our multilingual dataset integration approach. The SAFE Challenge evaluation strategy involved selecting the best-performing model architecture for each task based on the submitted model. WavLM Large model (iteration 4) proved optimal for Tasks 1 and 2. For Task 3 (laundered audio), MAE-AST Frame's more robust temporal modeling against adversarial processing made it the preferred choice, demonstrating the value of architectural diversity in handling different threat scenarios.

\subsection{Task-Specific Analysis}

Our evaluation reveals distinct challenges across the three SAFE Challenge tasks. Task 1 (unmodified audio) shows fundamental detection challenges with WavLM Large achieving strong performance (BA = 0.810) while MAE-AST Frame reaches 0.640. Task 2 (processed audio) demonstrates our highest performance levels (BA = 0.819 with WavLM Large), showing resilience to compression and resampling operations. Task 3 (laundered audio) presents the greatest challenge, where WavLM Large shows consistent degradation (0.478-0.518 range) compared to Tasks 1-2, while MAE-AST Frame maintains more stable performance across all tasks (0.597-0.640 range for iteration 2, 0.536-0.640 for iteration 4).

\begin{table}[h]
\centering
\caption{Balanced Accuracy on Task 1 for Pristine and Generated Sources (Iteration 4, WavLM Large). Bold values indicate balanced accuracy less than or equal to 0.60.}
\renewcommand{\arraystretch}{1.1}
\begin{tabular}{l c | l c | l c}
\hline
\textbf{Pristine 1} & \textbf{Acc.} & \textbf{Pristine 2} & \textbf{Acc.} & \textbf{Generated} & \textbf{Acc.} \\
\hline
MandPod1 & 0.87 & EngPod & 0.86 & elevenlabs & 0.64 \\
FleurGer & 0.87 & FleurEng & 0.86 & fish & 0.81 \\
VSPSemi & \textbf{0.58} & DigCass & 0.71 & hierspeech & 0.87 \\
YTPhone & 0.87 & Dipco & 0.84 & kokoro & 0.87 \\
VSPDoc & 0.84 & Librivox & 0.69 & parler & 0.86 \\
ArabCorpus & 0.87 & OldRadio & 0.69 & seamless & 0.76 \\
HQPod & 0.83 & PhoneHome & \textbf{0.53} & style & 0.86 \\
JapSWave & \textbf{0.39} & RussAudiobook & \textbf{0.52} & cartesia & \textbf{0.47} \\
Conf & \textbf{0.48} & MandPod2 & 0.86 & f5 & 0.63 \\
VSPHomeMic & 0.87 & RadioDrama & 0.81 & metavox & \textbf{0.58} \\
VSPProf & 0.84 & & & zonos & 0.65 \\
\hline
\end{tabular}
\vspace{0.2mm}

\footnotesize
\textit{Legend:}
MandPod1/2 = Mandarin Podcast 1/2; FleurGer/Eng = Fleurs German/English; VSPSemi/Prof = VSP Semi-professional/Professional;
YTPhone = YouTube phonecall; VSPDoc = VSP Documentary; HQPod = High Quality Podcasts; JapSWave = Japanese Shortwave
\label{tab:task1_sources}
\end{table}

\begin{table}[ht]
\centering
\caption{Balanced Accuracy for Task 2 (Processed) and Task 3 (Laundered) Sources (Iteration 4, WavLM Large). Bold values indicate balanced accuracy less than or equal to 0.60.}
\renewcommand{\arraystretch}{1.1}
\begin{tabular}{l c | l c | l c}
\hline
\multicolumn{2}{c|}{\textbf{Processed 1}} & \multicolumn{2}{c|}{\textbf{Processed 2}} & \multicolumn{2}{c}{\textbf{Laundered}} \\
\textbf{Source} & \textbf{Task 2} & \textbf{Source} & \textbf{Task 2} & \textbf{Source} & \textbf{Task 3} \\
\hline
aac 16k & 0.80 & pitch shift & 0.81 & car & \textbf{0.51} \\
encodec & 0.84 & resample $\downarrow$ & 0.77 & played & 0.67 \\
focalcodec & 0.83 & resample $\uparrow$ & 0.76 & reverb & 0.64 \\
mp3-aac-mp3 & 0.84 & sem-codec & 0.74 & all 3 & \textbf{0.49} \\
mp3-aac 16k & 0.81 & snac & 0.73 & & \\
mp3 16k & 0.79 & speech filt. & 0.79 & & \\
mp3 VBR & 0.79 & time stret. & 0.85 & & \\
noise & \textbf{0.52} & vorbis 16k & 0.75 & & \\
opus 16k & 0.73 & phone audio & 0.85 & & \\
\hline
\end{tabular}
\label{tab:tasks23_sources}
\end{table}

\subsection{Source-Level Performance Analysis} \label{source_level}

Source-level analysis was conducted for each major SAFE Challenge task using the Iteration 4 - WavLM Large system. The results for Task 1 (unmodified audio) are summarized in Table \ref{tab:task1_sources}, which lists balanced accuracy for both pristine and generated sources. Tasks 2 and 3 (processed audio and laundered audio, respectively) are reported in Table \ref{tab:tasks23_sources}. In these tables, bold values indicate cases where balanced accuracy was $\leq 0.60$, highlighting challenging sources for our system.

As seen in Table \ref{tab:task1_sources}, most pristine (real) sources exhibit high detection accuracy ($\geq0.80$), notably \textit{MandPod1}, \textit{FleurGer}, \textit{EngPod}, and \textit{VSPHomeMic}. However, several real sources stand out as considerably more challenging, with balanced accuracy of $\leq 0.60$: \textbf{VSPSemi} (\textbf{0.58}), \textbf{JapSWave} (\textbf{0.39}), \textbf{Conf} (\textbf{0.48}), \textbf{PhoneHome} (\textbf{0.53}), and \textbf{RussAudiobook} (\textbf{0.52}). This variability points to vulnerabilities in detecting authentic, diverse, or lower-quality real-world audio.

The generated sources (right-most column of Table \ref{tab:task1_sources}) generally achieve high detection rates as well. However, certain synthetic systems are more difficult to detect, including \textbf{cartesia} (\textbf{0.47}) and \textbf{metavox} (\textbf{0.58}). Most mainstream approaches, such as \textit{hierspeech}, \textit{kokoro}, and \textit{openai}, are robustly detected ($>0.80$), while the worst-performing synthetic and pristine sources may warrant special attention for system improvement.

Table \ref{tab:tasks23_sources} shows the balanced accuracy for each processed and laundered source. For processed sources (Task 2), the majority maintain balanced accuracy above 0.70, with a few notable exceptions: the \textbf{noise} condition (\textbf{0.52}) significantly lowers accuracy, showing that added noise can effectively degrade detection reliability. All other processed sources—various codecs, resampling, pitch shift, semantic encoding—sustain robust detection (typically $0.73-0.85$).

Task 3 (laundered audio) remains challenging, with the lowest performance: \textbf{car} (\textbf{0.51}) and \textbf{played reverb car} (\textbf{0.49}) both fall below the 0.60 threshold, and even the best-performing laundered source (\textit{played}, $0.67$) shows considerable performance degradation compared to unprocessed tasks.

\subsection{ITW Evaluation}

Performance on the In-The-Wild (ITW) benchmark provides crucial validation of our approach's generalization capabilities to real-world deepfakes from social media platforms. Our results demonstrate substantial improvements across iterations, with balanced accuracy progressing from 0.616 (baseline) to 0.905 (Iteration 4, WavLM Large) and EER reducing from 35.61\% to 8.42\%. Each iteration shows significant improvement: multi-dataset integration (BA: 0.616 → 0.875), audio length optimization (BA: 0.856), and strategic integration achieving optimal performance (BA: 0.905, EER: 8.42\%).

\section{Conclusion} \label{conc}
This work demonstrates that strategic integration of multilingual data sets significantly improves detection of deepfake audio in diverse scenarios. Our systematic approach, combining six complementary datasets with SSL-based architectures, achieved competitive SAFE Challenge rankings (2nd-3rd place) and substantial ITW benchmark improvements (EER: 35.61\% → 8.42\%).

Our key findings include: (1) dataset diversity provides dramatic performance gains over single-dataset approaches, (2) WavLM Large and MAE-AST Frame offer complementary strengths for different threat scenarios, (3) longer audio segments improve detection of processed artifacts, and (4) laundered audio presents critical challenges with false positive vulnerabilities that require careful deployment consideration.

Our source-level analysis reveals specific failure patterns and architectural trade-offs that inform future research directions. Although substantial progress has been made on clean and processed audio detection, adversarial laundering remains a significant challenge requiring continued investigation.



\bibliographystyle{IEEEtran}
\bibliography{references}

\end{document}